\documentclass[superscriptaddress,twocolumn,secnumarabic,amssymb, nobibnotes, aps, prd]{revtex4-1}

\usepackage{graphicx,array,multirow}
\usepackage{color}
\usepackage{amsmath}
\usepackage{ulem}

\newcommand{\lepton}{\ell}
\newcommand{\fshadron}{\textrm{X}}
\newcommand{\nucleus}{\textrm{A}}
\newcommand{\proton}{\textrm{p}}
\newcommand{\neutron}{\textrm{n}}

\newcommand{\pproton}{p_\proton}
\newcommand{\thetaproton}{\theta_\proton}
\newcommand{\pmu}{p_\mu}
\newcommand{\pvmu}{\vec{p}_\mu}
\newcommand{\pvnu}{\vec{p}_\nu}
\newcommand{\qv}{\vec{q}}

\newcommand{\thetamu}{\theta_\mu}

\newcommand{\dpt}{\delta p_\textrm{T}}
\newcommand{\ptl}{\vec{p}_\textrm{T}^{\,\mu}}
\newcommand{\ptlp}{\vec{p}_\textrm{T}^{\,\mu\,(\proton)}}
\newcommand{\ptni}{\vec{p}_\textrm{T}^{\,\proton}}

\newcommand{\genie}{\textsc{genie}}

\newcommand{\gevc}{\textrm{GeV}/\textit{c}}

\newcommand{\gev}{\textrm{GeV}}
\newcommand{\zeropi}{0\pi}
\newcommand{\onepi}{1\pi}

\newcommand{\alr}{A_\textrm{LR}}
\newcommand{\adler}{\phi^\ast}
\newcommand{\adlertheta}{\theta^\ast}
\newcommand{\adlerx}{\hat{x}^\ast}
\newcommand{\adlery}{\hat{y}^\ast}
\newcommand{\adlerz}{\hat{z}^\ast}
\newcommand{\ainput}{A_0}
\newcommand{\wa}{w_1}
\newcommand{\wb}{w_2}
\newcommand{\alrres}{A_\textrm{RES}}
\newcommand{\alrtot}{A_\textrm{All}}
\newcommand{\nres}{N_\textrm{RES}}
\newcommand{\ntot}{N_\textrm{All}}

\newcommand{\nresonepi}{\nres^{\onepi}}
\newcommand{\ntotonepi}{\ntot^{\onepi}}
\newcommand{\alrreszeropi}{\alrres^{\zeropi}}
\newcommand{\alrresonepi}{\alrres^{\onepi}}
\newcommand{\alrtotzeropi}{\alrtot^{\zeropi}}
\newcommand{\alrtotonepi}{\alrtot^{\onepi}}
\newcommand{\absof}{S_\pi}
\newcommand{\clr}{C_\textrm{LR}}
\newcommand{\fn}{f_N^{\onepi}}
\newcommand{\fa}{f_A}
\newcommand{\nucleon}{\textrm{N}}
\newcommand{\pvmus}{\vec{p}_\mu^{\,\ast}}
\newcommand{\pvnus}{\vec{p}_\nu^{\,\ast}}
\newcommand{\pnucleon}{\vec{p}_\nucleon^{\,\ast}}
\newcommand{\pvmusa}{\vec{p}_{\mu1}^{\,\ast}}
\newcommand{\pvnusa}{\vec{p}_{\nu1}^{\,\ast}}
\newcommand{\pnucleona}{\vec{p}_{\nucleon1}^{\,\ast}}

\newcommand{\iniN}{\mathcal{N}}
\newcommand{\nhmproton}{n^\nucleon}
\newcommand{\nhmprotonl}{\nhmproton_\textrm{L}}
\newcommand{\nhmprotonr}{\nhmproton_\textrm{R}}
\newcommand{\nhmpion}{n^\pi}

\newcommand{\ako}{A_\textrm{KO}}

\newcommand{\plotwidth}{0.48}

\DeclareMathOperator{\sgn}{sgn}

\begin{document}

\title{  Pion-proton correlation in neutrino  interactions on nuclei}%

\author{Tejin Cai}
\affiliation{University of Rochester, Rochester, New York 14627 USA}

\author{Xianguo Lu}
\email{Xianguo.Lu@physics.ox.ac.uk}
\affiliation{Department of Physics, University of Oxford, Oxford, OX1 3PU, United Kingdom}

\author{Daniel Ruterbories}
\affiliation{University of Rochester, Rochester, New York 14627 USA}

\date{\today}%

\newcommand{\edit}[1]{{#1}}
\newcommand{\editt}[1]{{#1}}

\newcommand{\edittt}[1]{{\color{green} #1}}

\begin{abstract}

In neutrino-nucleus interactions, a proton produced with a correlated pion might exhibit a left-right asymmetry relative to the lepton scattering plane even when the pion is absorbed. Absent in other proton production mechanisms, such an asymmetry measured in charged-current pionless production could reveal the details of the absorbed-pion events that are otherwise inaccessible.  In this study, we demonstrate the idea of using  final-state proton left-right asymmetries to quantify the absorbed-pion event fraction and underlying kinematics.  This technique might provide critical information that helps constrain all underlying channels in neutrino-nucleus interactions in the GeV regime. 

\end{abstract}

\maketitle

\section{Introduction}

In the GeV regime, neutrinos interact with nuclei via neutrino-nucleon quasi-elastic scattering (QE), resonant production (RES), and deeply inelastic scattering (DIS). These primary interactions are embedded in the nucleus, where nuclear effects can modify the event topology. For example, in interactions where no pion is produced outside the nucleus,  one could find contributions from both RES that is followed by pion absorption in the nucleus---a type of final-state interactions (FSIs)---and two-particle-two-hole ($2p2h$) excitation~\cite{Nieves:2011yp} besides QE. This admixture of underlying channels complicates the experimental studies of neutrino oscillations~\cite{ Abe:2015zbg, Acciarri:2015uup} where interaction cross sections and neutrino energy reconstruction are severely affected by nuclear dynamics.  

The recent development of data analysis techniques~\cite{Lu:2015tcr, Furmanski:2016wqo} allows the for separation of QE from non-QE processes in the aforementioned  zero-pion ($\zeropi$) topology. And yet, the remaining RES---referred to as \textit{absorbed-pion events} in this work---and $2p2h$ contributions occupy very similar phase space not only in the single-particle kinematics but also in the kinematic imbalance exploited by those techniques~\cite{Lu:2018stk, Abe:2018pwo}. Their experimental evidence is the otherwise unaccounted for measured excess of event rates~\cite{Rodrigues:2015hik, Gran:2018fxa}. Complementary to the  $\zeropi$ topology, pion production~\cite{Eberly:2014mra, Altinok:2017xua} has provided important constraints on both primary pion production and pion FSIs (cf. for example, Ref.~\cite{Mosel:2017ssx, Stowell:2019zsh}). Had there been no $2p2h$ contributions, details of absorbed-pion events could have been better determined. The lack of experimental signature to identify either process~\cite{Niewczas:2015iea, Weinstein:2016inx} is one of the biggest challenges in the study of neutrino interactions in the GeV regime. In this paper, we examine the phenomenon of pion-proton correlation and discuss the method of using final-state (i.e., post-FSI) protons to study absorbed-pion events. 

\section{Methodology}\label{sec:method}

\begin{figure}[!htp]
      \centering
      \includegraphics[width=0.4\textwidth]{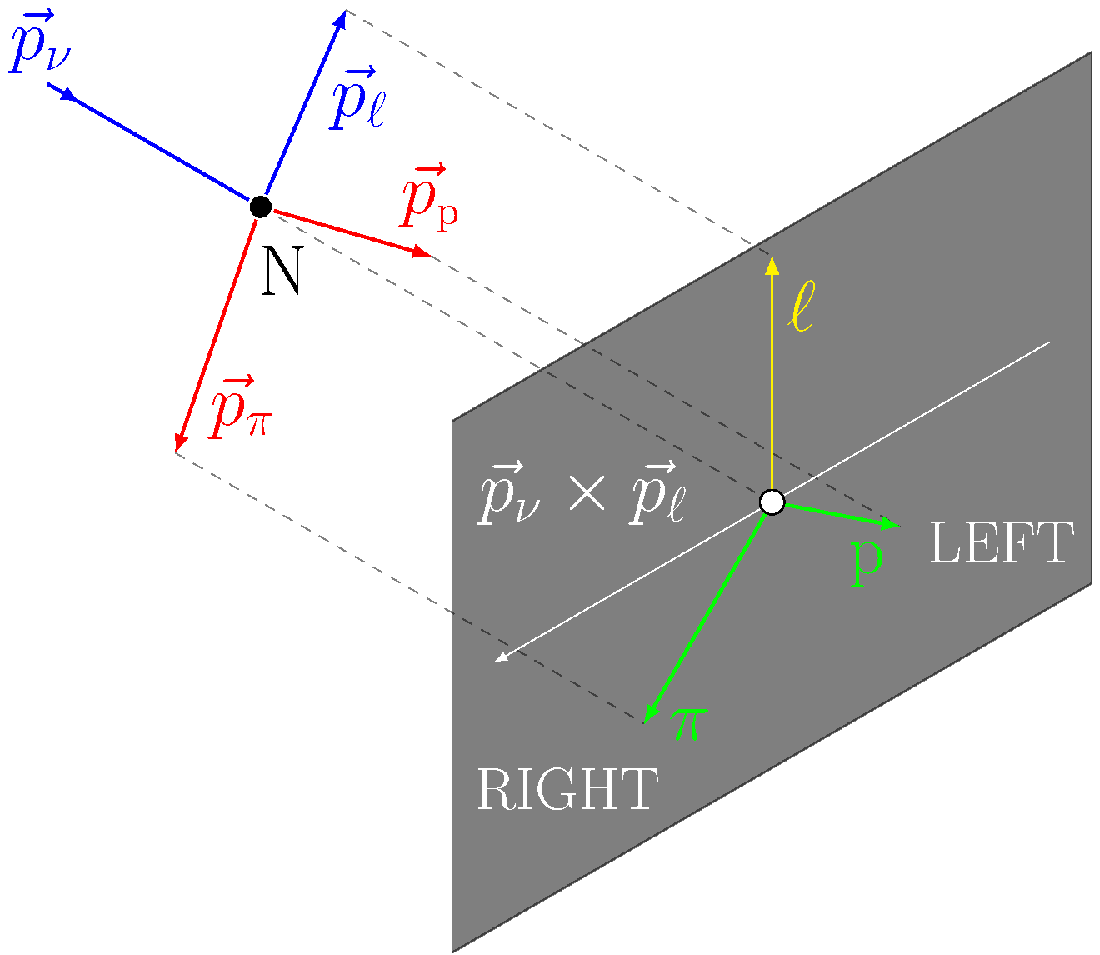}
      \caption{Schematic of the kinematics in reactions~(\ref{eq:nures}) and (\ref{eq:nupi0}). N is the target nucleon. The black screen is the transverse plane to the neutrino direction. Vectors on the screen represent the transverse projection of the respective momenta.}
      \label{fig:diagram}
  \end{figure}

In neutrino-nucleon scattering where a proton and a pion are produced,  
\begin{align}
\nu \proton &\to \lepton^- \proton \pi^+,\label{eq:nures}\\
\nu \neutron &\to \lepton^- \proton \pi^0,  \label{eq:nupi0}
\end{align}
with $\nu$, $\proton$, $\neutron$, $\lepton^-$, and $\pi$ being the neutrino, proton, neutron, charged lepton, and pions, respectively, the (lepton) scattering plane is defined as spanned by the momenta  of the  incoming  and outgoing leptons, $\vec{p}_{\nu}$ and $\vec{p}_{\lepton}$. With a stationary initial nucleon, momentum conservation requires that the final-state proton and pion occupy either side of the scattering plane. If we define the direction of $\vec{p}_\nu\times\vec{p}_\lepton$ to be the right of the plane, a proton on the left means a pion on the right, and a left-right asymmetry of the proton indicates an opposite asymmetry of the pion (Fig.~\ref{fig:diagram}). This is a spatial correlation between the pion and proton as a result of momentum conservation. In general neutrino interactions where a proton is produced, we define the proton left-right asymmetry (in the lab frame unless  otherwise stated) as
\begin{align}
    \alr\equiv\frac{N_\textrm{L}-N_\textrm{R}}{N_\textrm{L}+N_\textrm{R}},\label{eq:alr}
\end{align}
where $N_\textrm{L~(R)}$ stands for the event rate of the proton being on the left (right) of the scattering plane. The left-right asymmetries reported in existing measurements, as quoted in later discussions, are defined for the pion and hence need to be flipped (i.e. multiplied by $-1$) to translate to the corresponding proton asymmetries in reactions~(\ref{eq:nures}) and (\ref{eq:nupi0}).   

In neutrino interactions on nuclei where there are no pions in the final state, 
\begin{align}
\nu \nucleus &\to \lepton^- \proton \fshadron,\label{eq:nuAres}
\end{align}
with $\nucleus$ being the target nucleus and $\fshadron$ a hadronic system that doesn't include any pions, reactions~(\ref{eq:nures}) and (\ref{eq:nupi0}) take place on the constituent nucleons. The pion is subsequently absorbed in FSIs, while the proton propagates through the nucleus, carrying all primary information, including $\alr$. A detector then measures the final-state proton.  However, protons from QE and $2p2h$ are indistinguishable from these correlated protons. As a result, all measured primary information from reactions~(\ref{eq:nures}) and (\ref{eq:nupi0})  is diluted. The extent to which $\alr$ is reduced depends on the absorbed-pion fraction $\absof$ in the sample. In other words, by measuring the attenuation of the proton $\alr$, we could obtain $\absof$ in reaction~(\ref{eq:nuAres}).

This method relies on the following assumptions. First, there is a pion-proton correlation unique to pion production on free nucleons. In this work, we consider the left-right asymmetry with respect to the scattering plane. In the 1970s and 1980s, hydrogen and deuterium bubble chamber experiments provided the only measurements on (quasi-)free nucleon targets. With a deuterium target, ANL reported  $\alr=0.053\pm0.035$ of $\pi^+$ in  reaction~(\ref{eq:nures}) with a neutrino beam energy distribution that peaked near 0.9~GeV~\cite{Radecky:1981fn}.  In 2017, MINERvA  reported   $\alr=0.15\pm0.10$ of $\pi^0$ in the   $\proton\pi^0$ rest frame from reaction~(\ref{eq:nupi0})   on a hydrocarbon target at the Low-Energy NuMI beam with peak energy about 3 GeV~\cite{Altinok:2017xua}. These asymmetries indicate  the polarized production of resonances, dominated by $\Delta(1232)$, and are ascribed to the interference between the resonant and nonresonant amplitudes~\cite{Adler:1968tw, Adler:1975mt, Sobczyk:2018ghy, Kabirnezhad:2017jmf}. Most recently, T2K measured the left-right asymmetry in $\pi^+$ production on hydrocarbon over its off-axis neutrino beam spectrum peaking at 0.6~GeV  and reported $\alr=0.038\pm0.046$~\cite{Abe:2019arf}. It is consistent with a linear extrapolation of the previous measurements as a function of the beam energy. Currently, the bubble chamber data are dominantly constraining theory calculations on the nucleon-level asymmetry. It would no longer be sufficient because of  the required  precision and the potential beam-energy dependence for current and future experiments. A new method to  measure   pion-proton production on hydrogen has been proposed in Ref.~\cite{Lu:2015hea}, which would provide renewed  access to those asymmetries on free protons.  The second assumption is that,  the correlation needs to survive nuclear effects, such as Fermi motion and nucleon FSIs. In the following, we investigate the impact of nuclear effects on $\alr$ and demonstrate the idea of measuring absorbed-pion events with $\alr$ attenuation.

\section{Monte Carlo Calculation}

In this work, we use the neutrino event generator \genie~2.12~\cite{Andreopoulos:2015wxa} to study reaction~(\ref{eq:nuAres}) on carbon with a muon neutrino of energy 3~\gev.  In this simulation,  QE cross section is based on Ref.~\cite{LlewellynSmith:1971uhs},  RES is described by the Rein-Sehgal model~\cite{Rein:1980wg}, DIS by the  quark parton model (QPM)~\cite{Paschos:2001np} with the Bodek-Yang structure functions~\cite{Bodek:2002ps}, and  $2p2h$ by the Valencia model~\cite{Nieves:2011yp, Sobczyk:2012ms, Gran:2013kda, Schwehr:2016pvn}. PYTHIA6~\cite{Sjostrand:2006za} and models based on Koba-Nielsen-Olesen scaling~\cite{Koba:1972ng} are used to describe hadronization. The initial nuclear state is modeled as a local Fermi gas, and the FSI model is hA2015~\cite{Dytman:2011zz, Harewood:2019rzy}. 
A sample of charged-current (CC)  events is generated. The $\zeropi$ sample for this study is selected by requiring that there are no final-state  mesons, and the final-state muon and proton satisfy the following criteria:
\begin{align}
1.5~\gevc<&\pmu<10~\gevc,~\thetamu<20^\circ,\label{eq:pmuon}\\
0.45~\gevc<&\pproton,~\thetaproton<70^\circ,\label{eq:thetaproton}
\end{align}
where $\pmu$ and $\thetamu$ ($\pproton$ and $\thetaproton$) are the muon (proton) momentum and polar angle with respect to the neutrino direction, respectively.  These criteria are inspired by MINERvA's recent mesonless production measurement~\cite{Lu:2018stk}. In  reaction~(\ref{eq:nuAres}), because the final-state proton may be the knock-out product of a primary neutron undergoing FSI, an additional primary pion-neutron channel, 
\begin{align}
    \nu \neutron \to \lepton^- \neutron \pi^+,\label{eq:nunpiplus}
\end{align}
 is available in addition to the reactions~(\ref{eq:nures}) and~(\ref{eq:nupi0}).
 
In order to study $\alr$ attenuation in a controlled way, we removed the \genie~in-built empirical angular weight and restored isotropy in the resonant decay. The resulting distribution of the Adler angle $\adler$~\cite{Radecky:1981fn, Altinok:2017xua}---redefined here as the nucleon azimuth in the primary pion-nucleon  rest frame, cf. Appendix~\ref{sec:adler}---is flat, giving a zero intrinsic left-right asymmetry. We then introduce the left-right asymmetry by weighting $\adler$~\cite{Stowell:2016jfr}, where right is $0<\adler<180^\circ$. The following weighting schemes are used for comparison $(0^\circ\leq\adler<360^\circ)$:
\begin{align}
\wa(\adler)&=1+\sgn(\adler-180^\circ)\cdot\ainput,\label{eq:wa}\\
\wb(\adler)&=1-\frac{\pi}{2}\ainput\sin\adler,\label{eq:wb}
\end{align}
where $\sgn(x)$ is the sign function and $\ainput$ is the input $\alr$ in the primary pion-nucleon rest frame. A similar approach can be found in Ref.~\cite{Sanchez:2015yvw}. Because $N_\textrm{L,R}$ are integrals in the respective regions, their asymmetry is not sensitive to the modulation of the weights away from the left-right boundary, $\adler=0^\circ$ and $180^\circ$.  Therefore, the $\sgn$-and $\sin$-weighting schemes are general as they characterize different transitional behaviors---abrupt and smooth, respectively.   The  weight is applied to all RES channels before FSI. The post-FSI proton $\alr$ is then calculated for the leading (i.e. highest-momentum)  proton  for the absorbed-pion subsample and the overall sample, labeled by $\alrres$ and $\alrtot$, respectively. 

To characterize the impact of nuclear effects on the asymmetry, we organize the simulated events according to  the  transverse momentum imbalance~\cite{Lu:2015tcr, Lu:2018stk, Abe:2018pwo}, 
\begin{align}
    \dpt  &= \left|\ptl + \ptni\right| \label{eq:dpt},
\end{align}
where $\ptlp$ is the muon (proton) transverse momentum with respect to the neutrino direction. If there is no FSI, $\dpt$ is the transverse projection of the initial nucleon momentum due to Fermi motion; in full it represents the sum of the initial nucleon momentum and any intranuclear momentum exchange including FSI. Higher-order accuracy in describing the Fermi motion can be achieved by the 3-dimensional momentum imbalance introduced in Ref.~\cite{Furmanski:2016wqo} and measured by MINERvA~\cite{Lu:2018stk}. 

\begin{figure}[!htp]
      \centering
      \includegraphics[width=\plotwidth\textwidth]{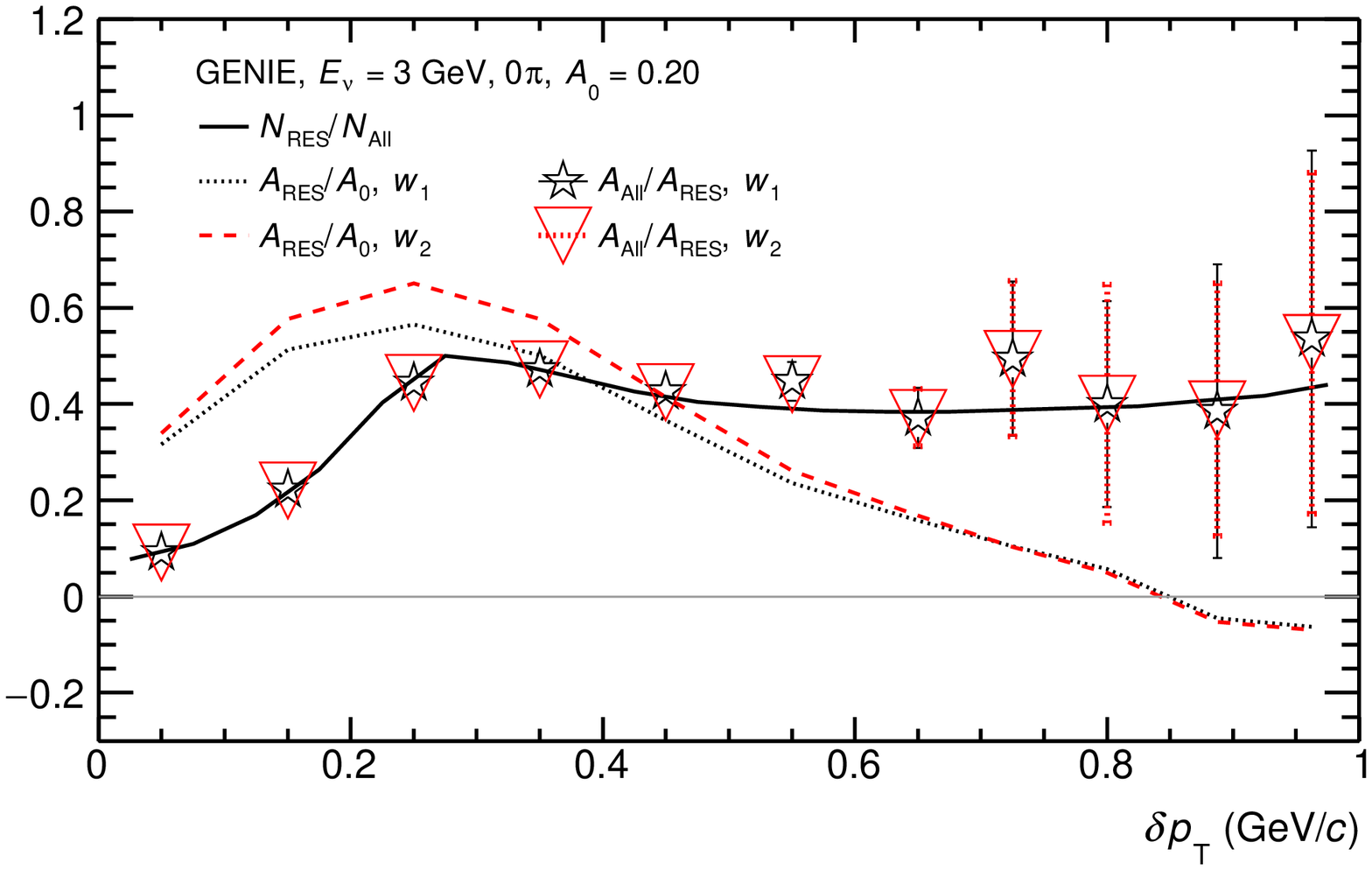}
      \caption{Absorbed-pion fraction $\absof$ ($=\nres/\ntot$) and asymmetry ratios as a function of $\dpt$ in the \genie~$\zeropi$ sample. The input asymmetry is $\ainput=0.20$. The asymmetry ratios between primary $\ainput$, and post-FSI $\alrres$ and $\alrtot$, are shown for the two weighting schemes $\wa$ and $\wb$.  The error bars represent statistical uncertainties in the simulation. The proposed experimental observable $\alrtot$ can be read out from the figure as  $\ainput\cdot(\alrres/\ainput)\cdot(\alrtot/\alrres)$, which is about $0.20\times0.5\times0.3=0.03$ at $\dpt\sim0.2$~\gevc.}
      \label{fig:zeropi}
  \end{figure}

The absorbed-pion fraction $\absof$ in the \genie~$\zeropi$ events is shown in Fig.~\ref{fig:zeropi}. It gradually increases with $\dpt$ to about 40\% and then forms a  plateau. Below $\dpt\sim0.3~\gevc$ is the Fermi-motion region dominated by QE. Above $\dpt\sim0.3~\gevc$ the absorbed-pion  and $2p2h$ events are the major contributions predicated with comparable size by \genie. Because the event rate difference [the numerator in Eq.~(\ref{eq:alr})] is the same in both the absorbed-pion and overall samples, the respective asymmetries, $\alrres$ and $\alrtot$, are inversely proportional to the sample size, $\nres$ and $\ntot$. Hence we have 
\begin{align}
    \absof=\nres/\ntot=\alrtot/\alrres,
\end{align}
as is demonstrated in Fig.~\ref{fig:zeropi} independent of the weighting scheme.

In experiments, while the overall asymmetry $\alrtot$ is accessible, $\alrres$ needs to be separately measured on free nucleons ($\ainput$) and folded with nuclear effects. In this work, the primary asymmetry is reduced by about 50\% ($\alrres/\ainput$ in Fig.~\ref{fig:zeropi}) in the Fermi-motion region due to the ``wobbling'' of the  primary pion-nucleon rest frame.  As $\dpt$ increases, the effect of nucleon FSI  is expected to increase, further reducing $\alrres$ due to nucleon-nucleon knockout. The asymmetry becomes zero at $\dpt\sim0.8~\gevc$ and then flips sign. The asymmetry-flip is the signature of knock-out protons from absorbed pions. Recall that the primary pion and nucleon have opposite asymmetries. An absorbed-pion knockout will carry part of  the pion asymmetry. If energetically more favourable, they will contribute a negative component to  $\alrreszeropi$. A quantitative analysis is presented in Section~\ref{sec:discussion}. We see that the impact from nuclear effects in the Fermi-motion region is smaller than the attenuation by uncorrelated protons characterized by $\alrtot/\alrres$.  A  measurement of $\absof$ with a model-dependent calculation of $\alrres$ would be interesting at $\dpt<0.4~\gevc$. This would already cover the whole onset region and reach the plateau. Furthermore, $\alrres$ could also be estimated in complementary pion production measurements as follows.

\begin{figure}[!htp]
      \centering
      \includegraphics[width=\plotwidth\textwidth]{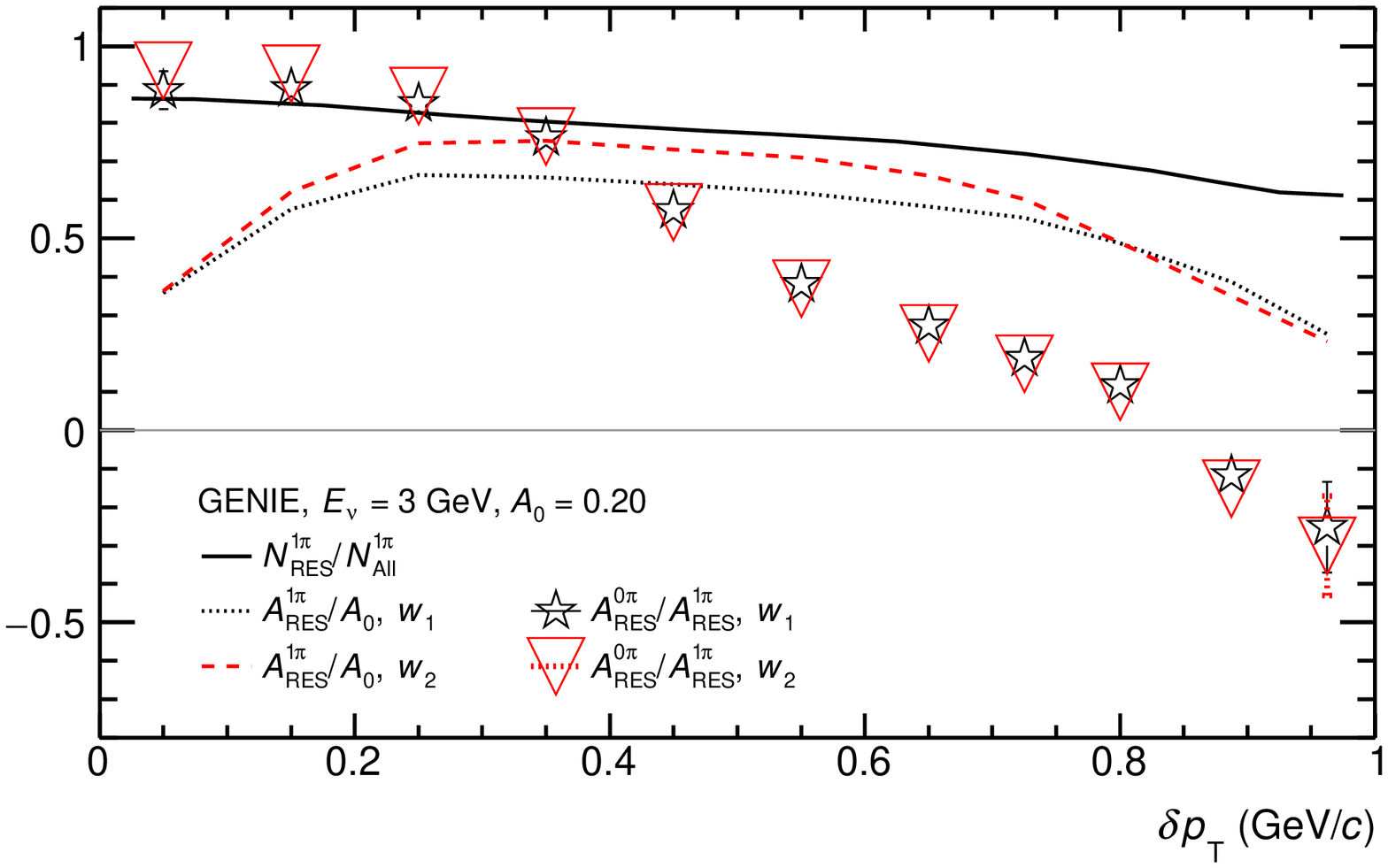}
      \caption{RES event fraction $\nresonepi/\ntotonepi$ and asymmetry ratios as a function of $\dpt$ in the \genie~$\onepi$ sample, where $\nresonepi$ and $\ntotonepi$ are the event rate of the $\onepi$ RES and overall sample, respectively.  The asymmetry ratios between primary $\ainput$, and the post-FSI $\alrreszeropi$ and $\alrresonepi$ in the respective $\zeropi$ and $\onepi$ sample, are shown for the two weighting schemes $\wa$ and $\wb$.}
      \label{fig:tagpi}
  \end{figure}

In addition to the $\zeropi$ sample, we simulate a one-pion ($\onepi$) sample by requiring exactly one final-state pion---and no other mesons---while keeping the same muon and proton selection. The pion ($\pi^{\pm,0}$) is only tagged but not used in the calculations, and the same analysis is done using only muon and proton kinematics using the same method used in the $\zeropi$ case.  As is shown in Fig.~\ref{fig:tagpi}, this sample is dominated by RES, especially at small $\dpt$, where the fraction is about 80\%; the rest is \genie-DIS events. Due to the absence of the absorbed-pion component, the proton asymmetry in this $\onepi$ sample, $\alrresonepi$,  is only reduced by nucleon FSI.  Overall, $\alrresonepi$ preserves $\ainput$ better than in the $\zeropi$ case, now called $\alrreszeropi$. Because to a good approximation the propagation of the primary nucleon and pion  do not interfere in the cold nuclear medium,  the difference between $\alrresonepi$ and  $\alrreszeropi$ is caused by the absorbed-pion knockout from the $\zeropi$ events.    As is indicated by  $\alrreszeropi/\alrresonepi$ in Fig.~\ref{fig:tagpi},  a ratio of about 80\% is observed  at small $\dpt$, which decreases and eventually becomes negative with increasing $\dpt$.

\begin{figure}[!htp]
      \centering
      \includegraphics[width=\plotwidth\textwidth]{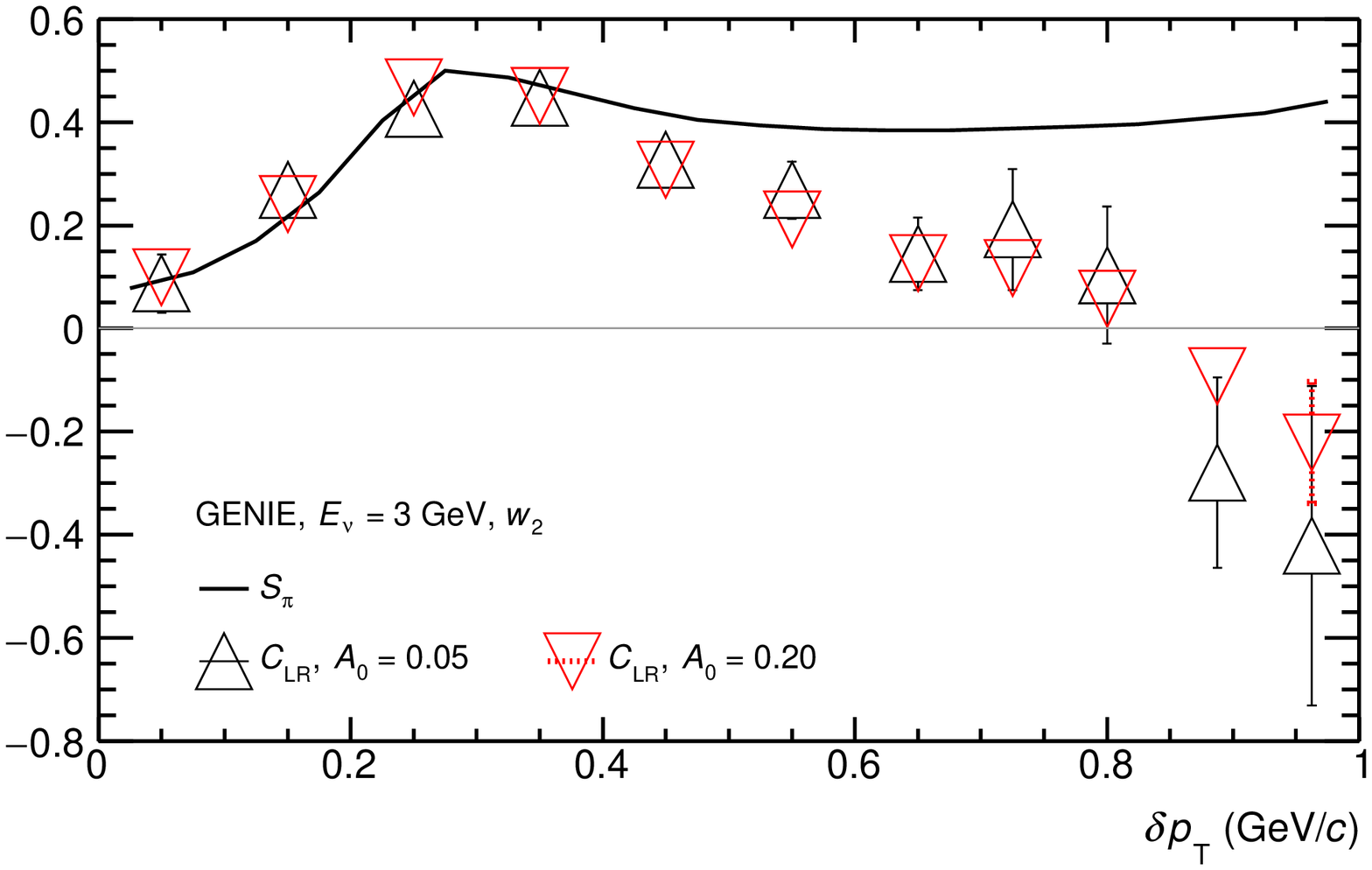}
      \caption{Absorbed-pion fraction $\absof$ in the \genie~$\zeropi$ sample compared to $\clr$ with different input asymmetries, $\ainput=$0.05 and 0.20. Weighting scheme  $\wb$ is used.}
      \label{fig:clr}
  \end{figure}

We can express the absorbed-pion fraction $\absof$  in $\zeropi$ events as  
\begin{align}
\absof&=\clr\frac{\fn}{\fa},\label{eq:spiclr}
\end{align}
with
\begin{align}
\clr&\equiv\frac{\alrtotzeropi}{\alrtotonepi},\\
\fn&\equiv\frac{\nresonepi}{\ntotonepi},\label{eq:fn}\\
\fa&\equiv\frac{\alrreszeropi}{\alrresonepi}, \label{eq:fa}   
\end{align}
where $\clr$ is experimentally accessible from  the $\zeropi$ and $\onepi$ samples, $\fn$ is the $\onepi$-event RES fraction correction, and $\fa$ is the knock-out correction. Because $\fa<1$ due to  absorbed-pion knockout in the $\zeropi$ sample and $\fn<1$ by definition, the two correction factors could be accidentally  canceling each other---this is indeed the case in \genie~with both being about~80\% at $\dpt<0.4~\gevc$. In Fig.~\ref{fig:clr} we see that $\clr$ alone describes $\absof$ rather accurately for the onset and up to the plateau. As a matter of fact, the  cancellation between $\fa$ and $\fn$ is not necessary for the proposed methodology. Experimental approaches to bring  both factors  under control are proposed in Section~\ref{sec:discussion}. Figure~\ref{fig:clr} further shows that the $\zeropi$ asymmetry-flip sets in at about $0.8~\gevc$. The variation of $\clr$ caused by different   $\ainput$  is shown to be statistical.   

\section{Discussion}\label{sec:discussion}

Comparing the two Adler angle $\adler$-weighting schemes [Eqs.~(\ref{eq:wa}) and~(\ref{eq:wb})], $\wa$ is more susceptible to the primary ``wobbling''  due to Fermi motion because of its abrupt transition at the left-right boundary ($\adler=0^\circ$ and $180^\circ$).   This explains the larger reduction of the asymmetry, namely smaller $\alrres/\ainput$, for $\wa$ in both Figs.~\ref{fig:zeropi} and~\ref{fig:tagpi}. Neither scheme has kinematic dependence.  In reality, however, the asymmetry is a function of  the primary nucleon and pion  kinematics. This requires precise phase-space matching (i.e. comparing the same final-state muon and proton phase space) between the $\zeropi$ and $\onepi$ samples such that the only difference between the two samples  is the pion FSI. Since pion FSI is decoupled from the primary pion-nucleon production, the phase-space matched asymmetry from the $\onepi$ sample can be used to infer the $\zeropi$ one.  In addition, the $\onepi$-tagging in this work does not have any phase space restriction. One would need to study the experimental threshold effect of the tagging techniques.

The effect of the absorbed-pion knockout in $\alrreszeropi$ can be quantified by the event rates of the leading protons, $\nhmproton$ and $\nhmpion$, originating from the primary nucleon ($\nucleon$) and pion, respectively. Assuming the asymmetries from these two components are exactly opposite, and they are of the same size as in $\alrresonepi$, we have
\begin{align}
    \alrreszeropi &\simeq\frac{\nhmproton-\nhmpion}{\nhmproton+\nhmpion}\frac{\nhmprotonl-\nhmprotonr}{\nhmproton},
\end{align}
where L and R denote the left and right parts of the event rates. The first factor on the RHS is the asymmetry between the two leading-proton origins, denoted as $\ako$, and the second one is identified as $\alrresonepi$ following the assumption. From  Eq.~(\ref{eq:fa}), we have
\begin{align}
\fa\simeq\ako.\label{eq:fako}
\end{align}
Therefore, the experimental observable $\clr$ reads
\begin{align}
    \clr\simeq\frac{1}{\fn}\absof\ako.
\end{align}
It has sensitivities to both the absorbed-pion event rate and kinematics. In addition to  the leading proton asymmetry in $\zeropi$ events,  one could consider the asymmetry of the subleading  proton. The corresponding $\ako$ will flip sign, as a leading proton from the primary nucleon means  a subleading proton from the primary pion. By combining the leading and subleading proton asymmetries in a two-proton $\zeropi$ sample, useful information on $\fa$ could be extracted. In the few GeV regime, the pion production and the corresponding pion-proton correlation in this \textit{de facto} shallow inelastic scattering (SIS) region is not well studied. This gives rise to significant uncertainties in   $\fn$. The SIS background could be reduced by restricting the invariant mass of the hadronic system, $W$, via calorimetry, like for example in Ref.~\cite{Le:2019jfy}, or by restricting the tagged pion momentum to reduce high-$W$ contributions.

The required sample size $\ntot$ of an asymmetry measurement depends on the asymmetry $\alrtot$ and the targeted relative statistical uncertainty $\varepsilon$:
\begin{align}
\ntot\simeq\frac{1}{\varepsilon^2\alrtot^2}\textrm{, for }\alrtot^{2}\ll1.    
\end{align}
Therefore, for a primary resonant asymmetry $\alrres$ of 0.05--0.2, a measurement of $\alrtot$  with 30\% (relative) statistical uncertainty would require an order of 10--100K $\zeropi$-events  in the several-GeV neutrino energy region. In the sub-GeV region  where the pion production is  close to the threshold, the smaller amount of pion absorption only allows for sensitivity to larger $\alrres$. On the other hand, if we measure proton $\alrtot$ as a function of the muon kinematics, the determination of left/right only requires angular  but not momentum measurement. This would greatly increase the sample size. This analysis strategy is of particular interest to non-magnetised trackers. For example, the MINERvA Low-Energy  $\zeropi$ measurement~\cite{Lu:2018stk} selects elastically scattered and contained protons to ensure the proton momentum-by-range precision. The sample size could increase by a factor of about 3  (roughly the ratio between total and elastic proton-nucleus cross sections)   if the proton momentum is not required, giving about 12K signal events. Furthermore, the MINERvA data using the NuMI Medium-Energy beam peaking at about 6~GeV~\cite{Valencia:2019mkf}  would provide about 10 times more  statistics than the Low-Energy data, enabling a sensitivity to $\alrres\simeq0.05$. Finally, the systematic uncertainties of measuring the asymmetry  would be better controlled than for a cross-section measurement because of the cancellation of the flux uncertainty which is currently at the 10\% level. 

\section{Summary and OUtlook}

In this work, we discuss the idea of pion-proton correlation and its application in the study of absorbed-pion events using final-state protons. Pion left-right asymmetries have been previously measured in neutrino CC production of overtly pion-proton final states, as an indication of  the $\Delta(1232)$-resonance $\adler$-polarisation. We demonstrate with simulations that the correlated proton asymmetry, $\alr$, can be observed in neutrino $\zeropi$  production on nuclei. 

Compared to electron-nucleus and pion-nucleus (cf., for example, Ref.~\cite{PinzonGuerra:2018rju}) scattering where the absorbed pion can be inferred to by the missing energy-momentum, there has been no direct evidence of pion absorption in neutrino $\zeropi$ events. Measuring the proton left-right asymmetry alone would be the first demonstration. This could possibly be done with   the existing T2K~\cite{Abe:2018pwo, Abe:2019bsl},   MicroBooNE~\cite{Acciarri:2016smi}, NOvA~\cite{Ayres:2004js}, and MINERvA~\cite{Lu:2018stk, Valencia:2019mkf} data. The \genie-predicted asymmetry-flip observed at $\dpt\sim0.8~\gevc$ is closely related to the kinematics of the absorbed pions and its knockout. The crossover appears to be a robust experimental observable as it is shown to be independent of the weighting scheme and the strength of the input asymmetry. The reproduction of a measured crossover would be an important benchmark for  nuclear-effect modeling.

We further introduce $\clr$, the asymmetry ratio between $\zeropi$ and $\onepi$ events, as an experimental probe for the absorbed-pion  fraction $\absof$ up to the \genie-predicted plateau at $\dpt\sim0.4~\gevc$. It would be interesting to investigate further how this could constrain $2p2h$ contributions, as it has been conjectured that the $\dpt$ region 0.3--0.4$~\gevc$ is where current $2p2h$ models have largest deficit~\cite{Lu:2018stk}.  The full potential of the proposed  techniques could be realised by the existing MINERvA Medium-Energy~\cite{Valencia:2019mkf} and the future SBND~\cite{Mcconkey:2018gpr} data sets. 

The pion-proton correlation considered in this work comes from pion production with the interference between resonant and nonresonant  amplitudes. This would be the dominant interaction  channel at the DUNE~\cite{Acciarri:2015uup} energy. However, similar correlation also exists in rare processes such as neutrino deeply virtual meson production~\cite{Kopeliovich:2012dr, Siddikov:2019ahb}. Finally, even though we have only discussed neutrino interactions,  the antineutrino counterparts,  
\begin{align}
     \bar{\nu} \proton &\to \lepton^+ \proton \pi^-,\label{eq:nubarres}\\
         \bar{\nu} \proton &\to \lepton^+ \neutron \pi^0,\label{eq:nubarpi0res}\\
     \bar{\nu} \neutron &\to \lepton^+ \neutron \pi^-,\label{eq:nubarneutronres}\\
     \bar{\nu} \nucleus &\to \lepton^+ \proton \fshadron,\label{eq:nubarAres}
\end{align}
can be studied analogously. By revealing the absorbed-pion fraction in antineutrino interactions, these additional measurements could be useful for the $CP$-violation search using neutrino and antineutrino  oscillations .

\begin{acknowledgments}

We  thank  Trung~Le, Federico~Sanchez, and Clarence~Wret for helpful discussions. We thank Richard~Gran and  Anthony~Mann additionally for their helpful suggestions on the manuscript.    We thank Kevin~McFarland  for extensive discussions that inspired this project. T.C. and D.R. are supported by DOE (USA) Grant No. DE-SC0008475.  X.L. is supported by  STFC (UK) Grant No. ST/S003533/1.  

\end{acknowledgments}

\appendix
\section{Definition of Adler angles}\label{sec:adler}
The kinematics of the primary pion production considered in this work [Eqs.~(\ref{eq:nures}),~(\ref{eq:nupi0}),~(\ref{eq:nunpiplus}), and~(\ref{eq:nubarres})--(\ref{eq:nubarneutronres})] can be described by the following two-step reaction:
\begin{align}
    \nu \iniN &\to \mu \Delta,\\
    \Delta &\to \nucleon \pi,\label{eq:decay}
\end{align}
where $\iniN$ is the in-coming nucleon, and $\Delta$ is the $\Delta$ resonance (or more generally the 4-momentum sum of the out-going nucleon $\nucleon$ and pion $\pi$). The angular distributions of $\nucleon$ and $\pi$ are determined by the decay [Eq.~(\ref{eq:decay})] kinematics in the  $\Delta$ rest frame (quantities in this frame are denoted by  ``$^\ast$"), where the Adler angles $\adlertheta$ and $\adler$ are defined: if we denote the usual  Cartesian basis vectors as $\adlerx$, $\adlery$, and $\adlerz$, and the nucleon  3-momentum $\pnucleon$, we have
\begin{align}
 \pnucleon\cdot\adlerx  &=  \left|\pnucleon\right|\sin\adlertheta\cos\adler, \label{eq:adlerx}\\
  \pnucleon\cdot\adlery  &=  \left|\pnucleon\right|\sin\adlertheta\sin\adler,\label{eq:adlery} \\
 \pnucleon\cdot\adlerz  &=  \left|\pnucleon\right|\cos\adlertheta.\label{eq:adlerz} 
\end{align}
Throughout this work, we use the angles of the nucleon, instead of the pion, unless otherwise specified. 

In the lab frame, if $\iniN$ is stationary, such as in the scattering on hydrogen, the $\Delta$ momentum is along the direction of the lepton momentum transfer,
\begin{align}
    \qv\equiv\pvnu-\pvmu,
\end{align}
where $\pvnu$ is the neutrino momentum. Therefore, in the $\Delta$ rest frame, $\adlerz$ and $\adlery$ are naturally defined by  $\pvnus-\pvmus$ and $\pvnus\times\pvmus$, respectively~\cite{Radecky:1981fn}. 
  
However, when $\iniN$ is subject to Fermi motion, as is the dominant case in current experimental situations, there are two   prescriptions to obtain the Adler angles:
\begin{itemize}
    \item[(1)] via a direct boost from the lab frame to the $\Delta$ rest frame which we denote by   the subscript ``1", standing for ``one-boost frame". Then $\adlerz_1$ and $\adlery_1$ are  defined by $\pvnusa-\pvmusa$ and $\pvnusa\times\pvmusa$, respectively. The Adler angles $\adlertheta_1$ and $\adler_1$  are calculated from Eqs.~(\ref{eq:adlerx})--(\ref{eq:adlerz}) with $\pnucleona$.
    \item[(2)] via a first boost from the lab frame to the $\iniN$ rest frame, followed by a second one to the  $\Delta$ rest frame which we denote by the subscript ``2" for  ``two-boost frame".   The corresponding basis vectors   are similarly obtained as before, and so are the Adler angles $\adlertheta_2$ and $\adler_2$.  
\end{itemize}
These two $\Delta$ rest frames are identical if the  Fermi motion  of $\iniN$ is collinear with $\qv$ in the lab frame; but in general,  the respective momenta of a given particle are different. At first glance, the two-boost frame seems the only correct one because  the Fermi motion  is removed by its first boost. As a matter of fact, the two frames only differ by a rotation, known as the Wigner rotation~\cite{Wigner:1939cj}, and since  angles are preserved under a rotation, we have  $\adler_1=\adler_2$ and $\adlertheta_1=\adlertheta_2$. 

\bibliographystyle{apsrev4-1}

\bibliography{bibliography}

\end{document}